\documentclass[prl,floatfix,twocolumn,showpacs]{revtex4}

\usepackage{amsmath}
\usepackage{amssymb}
\usepackage{graphicx}

\newcommand{\eg}{\emph{e.}$\,$\emph{g.}}
\newcommand{\ie}{\emph{i.}$\,$\emph{e.}}

\newcommand{\romc}{{\operatorname{c}}}

\newcommand{\VECe}{{\boldsymbol{e}}}
\newcommand{\VECn}{{\boldsymbol{n}}}
\newcommand{\VECr}{{\boldsymbol{r}}}

\begin{document}

\title{Twist-bend instability for toroidal DNA condensates}

\author{Igor M. Kuli\'c}
\author{Denis Andrienko}
\author{Markus Deserno}

\affiliation{Max Planck Institut f\"ur Polymerforschung,%
             Ackermannweg 10,%
             55128 Mainz,%
             Germany}

\date{\today}
\begin{abstract}
  We propose that semiflexible polymers in poor solvent collapse in
  two stages.  The first stage is the well known formation of a dense
  toroidal aggregate.  However, if the solvent is sufficiently poor,
  the condensate will undergo a second structural transition to a
  twisted entangled state, in which individual filaments lower their
  bending energy by additionally orbiting around the mean path along
  which they wind.  This ``topological ripening'' is consistent with
  known simulations and experimental results.  It connects and
  rationalizes various experimental observations ranging from strong
  DNA entanglement in viral capsids to the unusually short pitch of
  the cholesteric phase of DNA in sperm-heads.  We propose that
  topological ripening of DNA toroids could improve the efficiency and
  stability of gene delivery.
\end{abstract}


\maketitle


Single polymers collapse from a random coil conformation to a dense
state once the solvent gets sufficiently poor \cite{GrosbergKhokh}.
For a flexible chain the condition of minimal surface energy yields an
approximately spherical globule, but for semiflexible polymers the
situation is more complex \cite{ToroidFormation}: The local structure
of the dense phase then consists of essentially straight chains with a
basically parallel alignment, in order to minimize bending energy and
maximize density, respectively.  Such a state can be characterized by
a smooth field of tangent vectors, but in the spherical case this
field must have at least two energetically unfavorable defects on the
surface (you can't comb a sphere). However, for a \emph{torus} many
defect-free fields are possible. Indeed, DNA, the probably best
studied semiflexible polymer, readily forms beautiful nanotori after
adding any one of a variety of possible condensing agents (like
polyethylenglycol (PEG), multivalent counterions, or bundling
proteins) to a dilute solution of DNA chains \cite{Bloomfield}.  These
tori are surprisingly monodisperse, having a radius comparable to the
persistence length of DNA ($\approx\,50\,{\rm nm}$) basically
independent of the condensation method \cite{Bloomfield}.

Consider such a condensate, in which the chain is wound up like a
garden hose to form a torus with axial and tubular radii $r_1$ and
$r_2$, respectively.  Since $r_1$ is the average radius of curvature
of the chain, a simple scaling analysis balancing a bending energy
$A/r_1^2$ per unit length of polymer and a surface energy $\sigma$ per
unit area of the torus yields~\cite{ToroidFormation} $r_1 \sim
(\sigma/A)^{-2/5}V^{1/5}$ and $r_2 \sim (\sigma/A)^{1/5}V^{2/5}$,
where the chain volume $V\sim r_1r_2^2$ as well as the packing density
are assumed constant.  Hence, the aspect ratio $\xi = r_1/r_2 \sim
(\sigma/A)^{-3/5}V^{-1/5}$ shrinks if $\sigma$ or $V$ increase (\ie,
if the solvent gets poorer or the chain longer), and thus the torus
``fattens''.  In this case it is no longer justified to calculate the
bending energy with some average radius of curvature
$\langle\rho\rangle = r_1$.  In fact, since $\langle \rho^{-2}
\rangle \ge \langle\rho\rangle^{-2}$ (by virtue of Jensen's
inequality), the actual curvature energy should be larger.  However,
the same argument indicates that the condensate can \emph{lower} its
bending energy by redistributing curvature more evenly.  In this
letter we demonstrate that indeed below a critical aspect ratio
$\xi_\romc$ (or above a critical surface tension $\sigma_\romc$) the
system spontaneously relaxes bending energy by \emph{twisting} the
bundle of polymer strands.

Indirect indications of such a twisted state can be found in computer
simulations~\cite{Simulations}.  Analyzing Cryo-EM experiments on DNA
toroids~\cite{Cryo} Hud et al.\ proposed non-local DNA winding and
equidistribution of bending in toroids~\cite{Hud95}.  Besides
indications for the non-trivial \emph{local} organization of DNA in
toroids there are several lines of evidence for a global,
\emph{topological}, non-triviality coming from \emph{in vivo} studies:
Certain bacteriophages, whose DNA is (due to a genetic modification)
no longer attached to their nucleocapsid, display unusually strong
knotting of the genome \cite{PhageKnotting}.  And the chirality of the
highly confined sperm-chromatin is surprisingly pronounced, with a
pitch 10 times shorter than in vitro \cite{Livolant}.  These findings
make us wonder whether there is a connection to the topological
ripening we will now discuss.

Let us begin our quantitative analysis of the situation by neglecting
the connectivity of the chain.  More specifically, we will first
formulate a \emph{local} theory which is based on the above mentioned
nematic field $\VECn$ of unit tangent vectors \cite{deGennesMayer}.
The path of the actual polymer will later be recovered as an integral
curve of this flow field, and its global topological properties can
then be studied. The elastic energy $e$ per unit volume, describing
the deviation from perfectly parallel alignment, is the Frank-Oseen
free energy of a uniaxial nematic liquid crystal~\cite{deGPro95}
\begin{equation}
  e \; = \;
  \frac{K_1}{2}\big(\nabla\cdot\VECn\big)^2 +
  \frac{K_2}{2}\big(\VECn\cdot(\nabla\times\VECn)\big)^2 +
  \frac{K_3}{2}\big(\VECn\times(\nabla\times\VECn)\big)^2 \ ,
  \label{eq:Frank}
\end{equation}
where the three terms correspond to splay, twist, and bend
deformations, respectively. Assuming the condensate to behave like an
incompressible liquid the first term (splay) must vanish identically
in order to maintain a constant polymer density throughout the
condensate \cite{foot0}, while the other two terms divide the elastic
energy between themselves.  The total energy is of course the integral
of Eqn.~(\ref{eq:Frank}) over the torus volume.

It is convenient to discuss this situation in suitable toroidal
coordinates $\{r, \vartheta,
\varphi\}$, defined by
\begin{subequations}
\begin{eqnarray}
  x & = & (r_1-r\cos\vartheta)\cos\varphi \ , \\
  y & = & (r_1-r\cos\vartheta)\sin\varphi \ , \\
  z & = & r\sin\vartheta \ .
\end{eqnarray}
\end{subequations}
The nematic field is now represented as $\VECn = n_r\VECe_r +
n_\vartheta\VECe_\vartheta + n_\varphi\VECe_\varphi$ (where the
$\VECe_i=\partial_i\VECr/|\partial_i\VECr|$ are the toroidal unit
tangent vectors).  When inserting this into Eqn.~(\ref{eq:Frank}) and
integrating over the volume, we obtain the elastic energy as a
functional of $n_r(r, \vartheta, \varphi)$, $n_\vartheta(r, \vartheta,
\varphi)$, and $n_\varphi(r, \vartheta, \varphi)$.  At the surface of
the torus the boundary condition $n_r(r=r_2)=0$ must hold, and owing
to rotational symmetry we will henceforth make the (nontrivial but
very reasonable) assumption that none of the coordinate functions
depends on $\varphi$.  These steps reduce the task of finding the
optimal polymer winding to a two-dimensional variational problem.

\begin{figure}
  \includegraphics{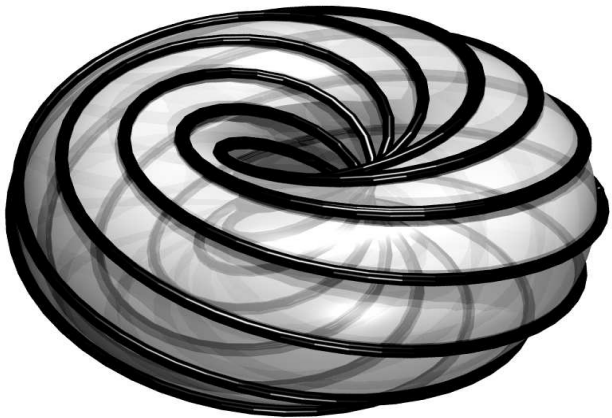}
  \caption{Illustration of the flow field on a toroidal condensate
  which features additional twist.  The aspect ratio is
  $\xi=1.5$.}\label{fig:torus}
\end{figure}

\begin{figure}
  \includegraphics[scale=0.85]{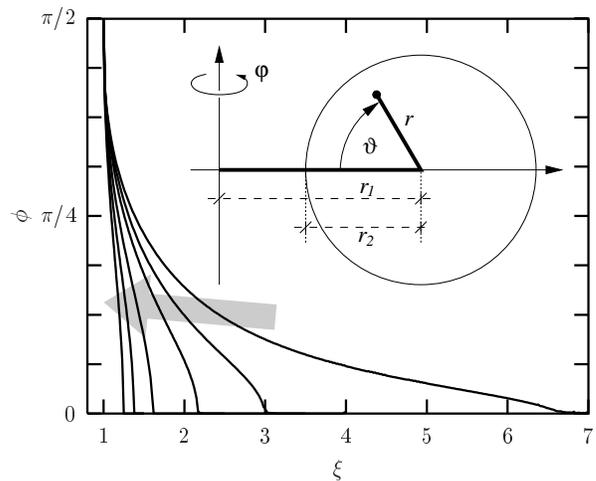}
  \caption{Order parameter $\phi:=\arccos(n_{\varphi,\min})$ as a
  function of the aspect ratio $\xi=r_1/r_2$.  The curves correspond
  to different ratios of elastic moduli, $\eta = K_2/K_3 \in \{0.01,
  0.05, 0.1, 0.2, 0.3, 0.4\}$, the gray arrow pointing toward
  increasing values.  The inset illustrates the toroidal coordinate
  system.}\label{fig:order}
\end{figure}

Now that the mathematical problem is formulated, let us first have a
look at the full solution, which we obtained numerically via a
conjugate gradient minimization \cite{PreFla92}.  The results confirm
the suspicions made above: For large enough aspect ratio $\xi=r_1/r_2$
the equilibrium nematic field is $\VECn = \VECe_\varphi$,
corresponding to simple circumferential winding of the polymer.  But
as the torus grows sufficiently fat, a continuous transition occurs to
a state in which (simultaneously throughout the entire torus) $\VECn$
acquires components in $\VECe_\vartheta$- and $\VECe_r$-direction,
\ie, the polymer additionally winds around the tubular circle, see
Fig.~\ref{fig:torus}.  This winding relaxes bending energy, but only
at the expense of the additional twist, which is zero when
$\VECn=\VECe_\varphi$.  Consequently, this twist instability occurs
more readily if the ratio $\eta=K_2/K_3$ of twist and bend modulus is
small.  All this is confirmed in Fig.~\ref{fig:order}, where we show
the maximum twist angle, which is a suitable order parameter, as a
function of the aspect ratio $\xi$ of the torus.  Note that the
calculation is reliable down to the value $\xi=1$, where the hole in
the torus degenerates to a point, the only possible tangent vector is
$\VECe_\vartheta$ and thus $\phi = \pi/2$.

The mathematical task of functional minimization can often be
accurately approximated by devising a variational ansatz which is
analytically tractable.  The following choice turns out to be
remarkably good: We will first assume that the nematic field does not
have a component in $\VECe_r$-direction.  The tangential boundary
condition $n_r(r=r_2)=0$ is then automatically taken care of.  The
remaining two components must satisfy the normalization condition
$n_\vartheta^2+n_\varphi^2=1$, and it thus suffices to specify one of
them, say $n_\vartheta$.  It is easy to check that any ansatz of the
form $n_\vartheta = f(r)/[1-(r/r_1)\cos\vartheta]$ with an arbitrary
function $f(r)$ yields a divergence free nematic field.  We choose a
linear $f(r)=\omega r/r_2$, \ie
\begin{equation}
  n_\vartheta(r,\vartheta;\omega)
  \; = \;
  \omega\,\frac{r/r_2}{1-(r/r_1)\cos\vartheta} \ ,
  \label{eq:ansatz}
\end{equation}
where $\omega$, which we may call the ``twisting strength'', is the
only free parameter of the ansatz.  With this choice we go back into
the Frank-Oseen free energy (\ref{eq:Frank}), calculate the
derivatives, and integrate over the volume of the torus.  Since the
sign of $\omega$ only determines the \emph{handedness} of the twisted
structure, it cannot influence the free energy $E$, which thus must be
an even function of $\omega$ \cite{foot1}.  An expansion in powers of
$\omega^2$ then yields
\begin{equation}
  \frac{E}{K_3\,r_2}
  \; = \;
  g_0(\xi) + g_2(\xi,\eta)\,\omega^2 + g_4(\xi,\eta)\,\omega^4 + \mathcal{O}(\omega^6) \ ,
  \label{eq:Landau_expansion}
\end{equation}
where the expansion coefficients $g_i$ are functions of the system
parameters \cite{foot2}. In particular, $g_0$ has the simple form
\begin{equation}
  g_0(\xi)
  \; = \;
  2\pi^2\Big(\xi-\sqrt{\xi^2-1}\Big)
  \; \stackrel{\xi\rightarrow\infty}{\longrightarrow} \;
  \pi^2/\xi \ .
\end{equation}
This term contributes even if the twist $\omega$ vanishes.  In fact,
it coincides with the bending energy of a chain of length $L$ which
has a curvature energy $A/\rho^2$ per unit length ($\rho$ being the
local radius of curvature), and which is wound without additional
looping within a torus of volume $V$---provided the (intuitively
clear) relation $K_3 V = A L$ holds.  This relates the nematic bending
modulus $K_3$ to the more usual polymer stiffness $A$.

While $g_0$ helped us to map our parameters, $g_2$ will localize the
phase transition.  The reason is that Eqn.~(\ref{eq:Landau_expansion})
has the form of a Landau free energy as it occurs for phase
transitions with a scalar order parameter, $\omega$, and it predicts a
continuous transition at the point where the coefficient of the
quadratic term vanishes, \ie, $g_2(\xi,\eta)=0$.  This results in the
phase boundary
\begin{equation}
  \eta \; = \;
  \frac{1}{2} +
  \frac{\xi^2-1}{4\,\xi^2}\bigg[1+6\,\xi\Big(\sqrt{\xi^2-1}-\xi\Big)\bigg] \ .
  \label{eq:ansatz_phase_boundary}
\end{equation}

\begin{figure}
  \includegraphics[scale=0.85]{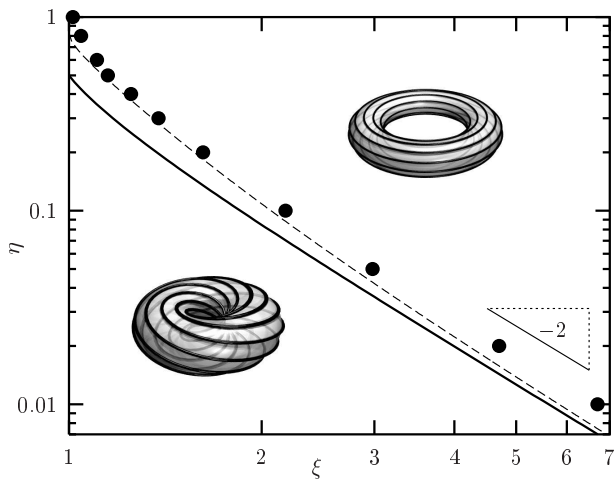}
  \caption{Structural phase diagram of the toroidally wound complex on
    a log-log scale.  For small $\xi=r_1/r_2$ and $\eta=K_2/K_3$ the
    polymer is wound in a twisted way, for large $\xi$ and $\eta$ it
    prefers to wind straight.  The dots are results from the full
    numerical minimization, the solid line is
    Eqn.~(\ref{eq:ansatz_phase_boundary}), and the dashed line stems
    from the ``improved ansatz'' (see text).}\label{fig:phasediagram}
\end{figure}

This boundary is shown in Fig.~\ref{fig:phasediagram}, together with
exact points originating from the full functional minimization.  It
quantifies the conclusion that toroidal condensates will spontaneously
twist if the aspect ratio $\xi=r_1/r_2$ is small (and the torus thus
fat), and if the ratio of elastic moduli $\eta=K_2/K_3$ is small, \ie\
if twisting a bundle is easy compared to bending it.  For large $\xi$
Eqn.~(\ref{eq:ansatz_phase_boundary}) asymptotically behaves like
$\eta \sim \frac{5}{16}\xi^{-2}$, thereby explaining the exponent
$-2$, which is also indicated in Fig.~\ref{fig:phasediagram} and which
actually describes the whole phase boundary quite well.  The agreement
between the simple one-parameter variational ansatz and the full
calculation is remarkably good.  It can even be improved by including
an additional prefactor $1-\frac{r}{r_1}$ into Eqn.~(\ref{eq:ansatz}).
Particularly for small $\xi$ this improved ansatz agrees better with
the exact answer, basically since the new prefactor cancels the
unphysical divergence of the denominator for $r\rightarrow r_1$ at
$\vartheta=0$.  The analytical expression for the phase boundary is
quite involved and will not be shown here, but it has the same large
$\xi$ asymptotics (see Fig.~\ref{fig:phasediagram}).

Both the ansatz as well as the full numerical solution point to an
upper critical ratio of elastic moduli, $\eta_\romc$, beyond which the
system will no longer spontaneously twist.  Even at the lowest
possible aspect ratio $\xi=1$ the energetic cost for twisting has
become so large that it no longer admits a bending relaxation.  The
ansatz (\ref{eq:ansatz}) gives $\eta_\romc = \frac{1}{2}$, the
improved ansatz gives $\eta_\romc \approx 0.829$, while the full
numerical solution suggests the deceptively simple result $\eta_\romc
= 1$.  We have no analytical support for the latter, but we also want
to remind that the limit $\xi \rightarrow 1$ is somewhat academic,
because our tacit assumption that the condensate shape is strictly
toroidal will most likely break down in this case \cite{Odijk}.

We have thus seen that the ratio $\eta$ and the torus geometry $\xi$
uniquely specify the twist-state of the condensate.  However, while
one can easily measure $\xi$ in an experiment (just by visual
inspection), it is hard to specify in advance.  In contrast to that,
the surface tension $\sigma$ can be readily changed (for instance via
the concentration of condensing agents), but its actual value is hard
to measure.  In our simple model it is of course not difficult to add
a tension term $\sigma$ times the torus surface $S$ to the condensate
energy.  Using $S\propto\xi^{1/3}V^{2/3}$, one can re-express the
twisting transition as being driven by increasing $\sigma$, and it
remains continuous.  However, practical considerations would advise to
tune $\sigma$ only for the purpose of modifying $\xi$, but
subsequently use $\xi$ as the independent variable.  This way one
needs no longer (neither theoretically nor practically) worry about
how a particular concentration of condensing agents gives rise to a
particular torus geometry.

After these local considerations it is time to study global aspects of
the polymer structure.  Let us start with the flow itself.  It can be
shown that incompressibility, $\nabla\cdot\VECn=0$, together with
axial symmetry causes the flow to be Hamiltonian---hence one more
conservation law exists~\cite{foot4}.  In our case it forces the flow
lines to stay on two-dimensional slightly deformed toroidal surfaces,
such that the total flow consists of a nested structure of invariant
tori.  In fact, our ansatz (\ref{eq:ansatz}) follows readily from the
quadratic Hamiltonian $H = \frac{1}{2}\xi\omega r^2$, which is
constant on circular tubular layers.  Of course, the actual polymer
has to switch between these layers, reminding us that irrespective of
twist none of the above structures can be realized without localized
defects \cite{Park98}.

There is one global aspect of the polymer structure in which it
differs fundamentally from a plainly wound torus: As is visible in
Fig.~\ref{fig:torus}, the path of the polymer threads it repeatedly
through the middle hole.  Moreover, the amount of this looping (as
measured \eg\ by the average change in $\vartheta$ per turn) depends
on the layer.  This effect implies that the entire polymeric strand is
heavily entangled with itself.  A rough estimate for $\xi=1.5$ and
polymer length $L=15 \mu{\rm m}$ gives about 30 threadings through the
hole.  In fact, were it not for the two free ends, these knotted
states were topologically inaccessible.  In other words, unlike the
initial collapse, the second stage, the structural ripening, relies on
the motion of the free chain ends and is thus a much slower process.

On the other hand this structural ripening meets no kinetic barriers
during the relaxation to its twisted ground state, downhill the free
energy landscape.  The motion of the two free ends is then
energetically directed and their local rearrangement does not involve
the highly improbable threading through the toroid hole in 3D space.
In addition, the weak chiral interaction of DNA molecules
\cite{Livolant} neglected above, gives rise to a (small) chiral term
in the elastic free energy, which might contribute to the symmetry
breaking and ``guide'' the twisting in a preferred direction.

It is also tempting to explain the unusually short DNA cholesteric
pitch (10 times shorter than \emph{in vitro}) in sperm chromatin
assuming that the weak DNA chirality merely determines the handedness
of twist whereas its pitch is given by the twisted state of polymer
strands after the topological ripening took place.  Moreover, the
severe knotting in highly dense phage heads, where the $2 {\rm nm}$
thick DNA has little space left for usual entropic entanglement
effects, suggests the same explanation---in particular since it is far
more pronounced for genetically modified phages in which the DNA is no
longer attached to the capsid and can thus undergo structural
ripening.

Finally, is the predicted effect strong enough to be of some relevance
for DNA condensation and gene delivery?  For typical experimental
parameters of DNA length $L = 15-30 \mu{\rm m}$, $\xi \sim 1.5-2.5$,
bending stiffness $A \sim 50 {k}_{\rm B}T{\rm\cdot nm}$, and
inter-helical distance $d \sim 3 \rm nm$ we obtain first the elastic
constant $K_{3} \sim A/d^{2}\sim 20 \rm pN$, which is dominated by the
bending stiffness \cite{deGennesMayer}.  The twist constant $K_{2}$
can be estimated by the decondensation force $\sim 2
\rm pN$ obtained in single molecule experiments with condenser
spermidine \cite{condenser_spermidine}.  The difference in elastic
energy between the twisted and untwisted states, as bounded below by
the the variational ansatz, lies in the range $15-30 k_{\rm B} T$.
This indicates that topological ripening stabilizes the condensate.
In addition, if the solvent quality abruptly improves, the twisted DNA
will unfold more inertly than its untwisted counterpart, due to heavy
entanglement with itself.  If this stabilization occurs on the typical
time scales relevant for gene therapeutical applications, it might
prevent a premature digestion of the genetic material by the host
organism and influence (positively) the efficiency of the gene
delivery process.


We thank H.~Schiessel, K.~Kremer, I.~Pasichnyk, and A.~Ryskin for helpful
discussions.




\begin{thebibliography}{99}
\bibitem{foot1} Here we first neglect the usually weak cholesteric
  interaction~\cite{Livolant} stemming from DNA chirality.  An explicit
  inclusion of a small chiral term would merely enhance the basic
  effect proposed above.
\bibitem{foot2} For the particular ansatz (\ref{eq:ansatz}) the
  resulting coefficient $g_4$ does in fact not depend on $\eta$; but
  generally it does.
\bibitem{foot4} The following transformation of the field and
  coordinates given in a cylindrical coordinate system, $\{\rho,z\}
  \rightarrow \{\rho^2/2,z\}$, $\{n_\rho,n_z\} \rightarrow \{\rho
  n_\rho, n_z\}$, together with the axial symmetry, reduces the 3D
  divergence-free nematic field to the (new) 2D divergence-free field.
  Then the director field can be written as $n_\rho = \rho^{-1}
  \partial H / \partial z$, $n_z = - \rho^{-1} \partial H / \partial
  \rho$, where $H(\rho,z)$ is the Hamiltonian.
\end{thebibliography}
\end{document}